\documentclass[fleqn,usenatbib]{mnras}

\usepackage[T1]{fontenc}

\DeclareRobustCommand{\VAN}[3]{#2}
\let\VANthebibliography\thebibliography
\def\thebibliography{\DeclareRobustCommand{\VAN}[3]{##3}\VANthebibliography}

\usepackage{graphicx}	
\usepackage{amsmath}	
\usepackage{amssymb}	
\usepackage{newtxtext,newtxmath}
\usepackage{color}

\title[Optical circular polarization of 3C 279 \& PKS 1510$-$089]{Constraints on magnetic field and particle content in blazar jets through optical circular polarization}

\author[I. Liodakis]{
I. Liodakis$^{1}$\thanks{E-mail: yannis.liodakis@utu.fi}, 
D. Blinov$^{2,3,4}$, S. B. Potter$^{5,6}$, and F. M. Rieger$^{7,8}$
\\
$^{1}$ Finnish Center for Astronomy with ESO, University of Turku, Quantum, Vesilinnantie 5, FI-20014, Finland\\
$^{2}$ Institute of Astrophysics, FORTH, GR-71110 Heraklion, Greece\\
$^{3}$ Department of Physics, University of Crete, 70013, Heraklion, Greece\\
$^{4}$ St. Petersburg State University, Universitetsky pr. 28, Petrodvoretz, 198504 St. Petersburg, Russia\\
$^{5}$ South African Astronomical Observatory, PO Box 9, Observatory, 7935, Cape Town, South Africa\\
$^{6}$ Department of Physics, University of Johannesburg, PO Box 524, Auckland Park 2006, South Africa\\
$^{7}$ Institute for Theoretical Physics (ITP), University of Heidelberg, Philosophenweg 12, 69120 Heidelberg, Germany\\
$^{8}$Max-Planck-Institut f\"{u}r Kernphysik, P.O. Box 103980, 69029 Heidelberg, Germany
}

\pubyear{2021}

\begin{document}
\label{firstpage}
\pagerange{\pageref{firstpage}--\pageref{lastpage}}
\maketitle

\begin{abstract}
Polarization offers a unique view in the physical processes of astrophysical jets. We report on optical circular polarization observations of two famous blazars, namely 3C~279 and PKS~1510$-$089, at high linearly polarized states. This is the first time PKS~1510$-$089 is observed in optical circular polarization. While only upper limits can be extracted from our observing campaign, the non-detection of optical circular polarization allows us to provide meaningful constraints on their magnetic field strength and jet composition. We find that high-energy emission models requiring high magnetic field strength and a low positron fraction can be excluded.

\end{abstract}




\section{Introduction}

Polarization is a powerful tool to understand the magnetic field structure and evolution in black hole jets \citep{Blandford2019,Hovatta2019}. Here we focus on a particular class of jets from supermassive black holes that are oriented towards the observer called blazars. Due to the synchrotron nature of their radiation, blazars are highly polarized and show variable, and often quite peculiar, polarization behavior. This behavior is connected to their high-energy emission \citep{Blinov2018}, the origin of which is still a mystery. The main mechanisms invoked to explain the high-energy emission in blazar jets are inverse-Compton scattering of relativistic electrons (also known as leptonic processes) by either internal (SSC) or external (EC) photon fields, and proton synchrotron, Bethe-Heitler pair production, pion decay etc. (also known as hadronic processes). Typically the distinction between these processes is investigated through spectral energy distribution modeling (SED, e.g., \citealp{Boettcher2013,Cerruti2020}). The Imaging X-ray Polarimetry Explorer (IXPE) will soon provide a new avenue of probing the origin of the high-energy emission in blazars through X-ray polarization \citep{Liodakis2019}. Here we explore a third direction, that of optical circular polarization.

The optical linear polarization degree in blazars can vary  between 0 and $>$30\% within just a few days (e.g., \citealp{Kiehlmann2016,Liodakis2020-II,Kiehlmann2021}). A fraction of them shows rotations of the optical polarization plane that are most often accompanied by  bright outbursts across the electromagnetic spectrum \citep{Marscher2008,Marscher2010,Bottacini2016,Raiteri2017,Uemura2017,Blinov2021}. While their optical and radio linear polarization (LP) properties have been getting plenty of attention lately \citep[e.g.,][]{Kovalev2020-II,Kravchenko2020,Blinov2021-II,Eventhorizon2021,Eventhorizon2021-II}, their optical circular polarization (CP) properties are unexplored. This is partly because of the scarcity of instruments capable of delivering CP measurements, but also because CP is expected to be low even in optimistic scenarios, requiring very sensitive instruments. Non-intrinsic CP can arise due to inverse-Compton scattering of low-energy radio photons, propagation effects, coherent radiation, accretion disk emission etc. (see \citealp{Rieger2005} for an overview).  Measurements of intrinsic CP, however, are particularly important because if the jet is made up of a pure electron-positron plasma then the expected intrinsic degree of CP is zero. Any confident detection of CP would (assuming overall charge-neutrality) imply at least a small fraction of protons in the jet. Therefore, CP observations offer a unique view on jet composition, which is otherwise not accessible. This, of course, has implications for the origin of the high-energy emission, especially in light of recent results on possible astrophysical neutrino associations with blazar jets \citep{IceCube2018,Plavin2020,Hovatta2021}.  A few attempts have been made to measure CP in blazars (e.g., \citealp{Valtaoja1993,Wagner2001,Hutsemekers2010}), however, there has not been a 3$\sigma$ detection of optical CP in blazars to this day. \cite{Wagner2001} reported a $<$2\% 3$\sigma$ detection of CP in 3C~279 using the VLT, however, those observations suffered from linear-to-circular polarization cross-talk which was not accounted for at the time \citep{Bagnulo2011}.

CP of a few percent has been detected in several blazars at radio wavelengths \citep[e.g.,][]{Homan2001-II,Thum2018,Myserlis2018,Hovatta2019-II}, however, it is most often attributed to Faraday conversion (which has a negligible effect in the optical bands) of the linear polarization due to intervening magnetized plasma \citep{MacDonald2018}. Therefore, simultaneous measurements of CP in both radio and optical, can help constrain the properties of the Faraday screen. In this letter, we present optical CP observations of two well-studied, and highly polarized blazars namely 3C~279 and PKS~1510$-$089. Assuming an intrinsic CP signal exists, we use the detected upper limits in combination with their linear polarization degree to constrain their magnetic field strength and jet composition.

\section{Observations \& data reduction}

\begin{table*}
\setlength{\tabcolsep}{11pt}
\centering
  \caption{Summary of observations for 3C~279 and PKS~1510$-$089. The columns are (1) name, (2) date of observations, (3) observing band, (4) linear polarization degree (\%), (5) polarization angle (degrees), and (6) circular polarization degree (\%). The uncertainties in columns 4, 5, and 6 correspond to the standard deviation of each measurement. }
  \label{tab:Table1}
\begin{tabular}{@{}cccccc@{}}
 \hline
 Name  & Date & Band & $\Pi_{l}$  & PA & $\Pi_{c}$  \\
  \hline
PKS~1510$-$089 & 59312.645 & {\it R} & 7.07 $\pm$ 0.51 & 5.0 $\pm$ 1.7 & -0.52 $\pm$ 0.37 \\
PKS~1510$-$089 & 59313.645 & {\it R} & 9.44 $\pm$ 1.45 & 4.6 $\pm$ 3.6 & 0.70 $\pm$ 1.02 \\
PKS~1510$-$089 & 59315.631 & {\it R} & 3.65 $\pm$ 1.06 & 65.0 $\pm$ 11.7 & -0.52 $\pm$ 0.76 \\
3C~279 & 59316.519 & {\it R} & 31.46 $\pm$ 0.42 & 126.4 $\pm$ 0.3 & -0.04 $\pm$ 0.30 \\
3C~279 & 59316.526 & {\it B} & 32.72 $\pm$ 1.01 & 127.5 $\pm$ 0.6 & -0.81 $\pm$ 0.73 \\
\hline
\end{tabular}
\end{table*}

3C~279 and PKS~1510$-$089 are famous for their bright outbursts, high variable emission, extreme polarization behavior as well as potentially multimessenger emission \citep{Marscher2010,Abdo2010-III,Fermi2016,Kreter2020,Blinov2021}.  The sources were observed during the nights of the 7, 8, 10 and 11th of April 2021,  using the HIgh-speed Photo-POlarimeter(HIPPO; \citealp{Potter2010}) on the 1.9-m telescope of the South African Astronomical Observatory.  HIPPO measures polarization using two contra-rotating 1/2 and 1/4 wave-plates, rotating at 10Hz and hence modulating the ordinary and extraordinary beams. The rotation speed is sufficient to effectively negate any errors due to atmospheric variations. Following \cite{Serkowski1974} we estimate the amplitude and phases of the modulations using a least-squares fit algorithm to fit the 4th and 8th harmonics (linear polarization) and the 6th harmonic \cite[circular polarization,][]{Potter2008,Potter2010}. Correction and efficiency factors are used to account for the fact that the modulated signal consists of a finite number of bins, instrumental polarization, position angle offsets, and a minor wavelength dependence due to retardance of the wave-plates. We estimated those factors by observing several polarized and non-polarized standard stars \citep{1982ApJ...262..732H,1988AJ.....95..900B}. Background sky polarization measurements were taken immediately preceding every observation. In order to not introduce a random non-zero circular polarization offset during sky subtraction, the circular polarized sky measurement was zeroed and only the unpolarized DC level was subtracted. The instrument is optimized for on-axis point source observations, hence it is not susceptible to linear-to-circular polarization cross-talk. Table \ref{tab:Table1} lists the linear and circular polarization parameters for both sources.  Our results confirm the sources as high LP blazars and provide CP constraints at the level of $\sim$1\%. Compared to archival observations, we find PKS1510$-$089's LP to be higher than average and 3C~279 to be at a very high LP state, near historical maximum \citep{Blinov2021,Blinov2021-II}.

\section{Constraints on the magnetic field strength and jet composition}

\begin{figure*}
\centering
\resizebox{\hsize}{!} {\includegraphics[width=\hsize]{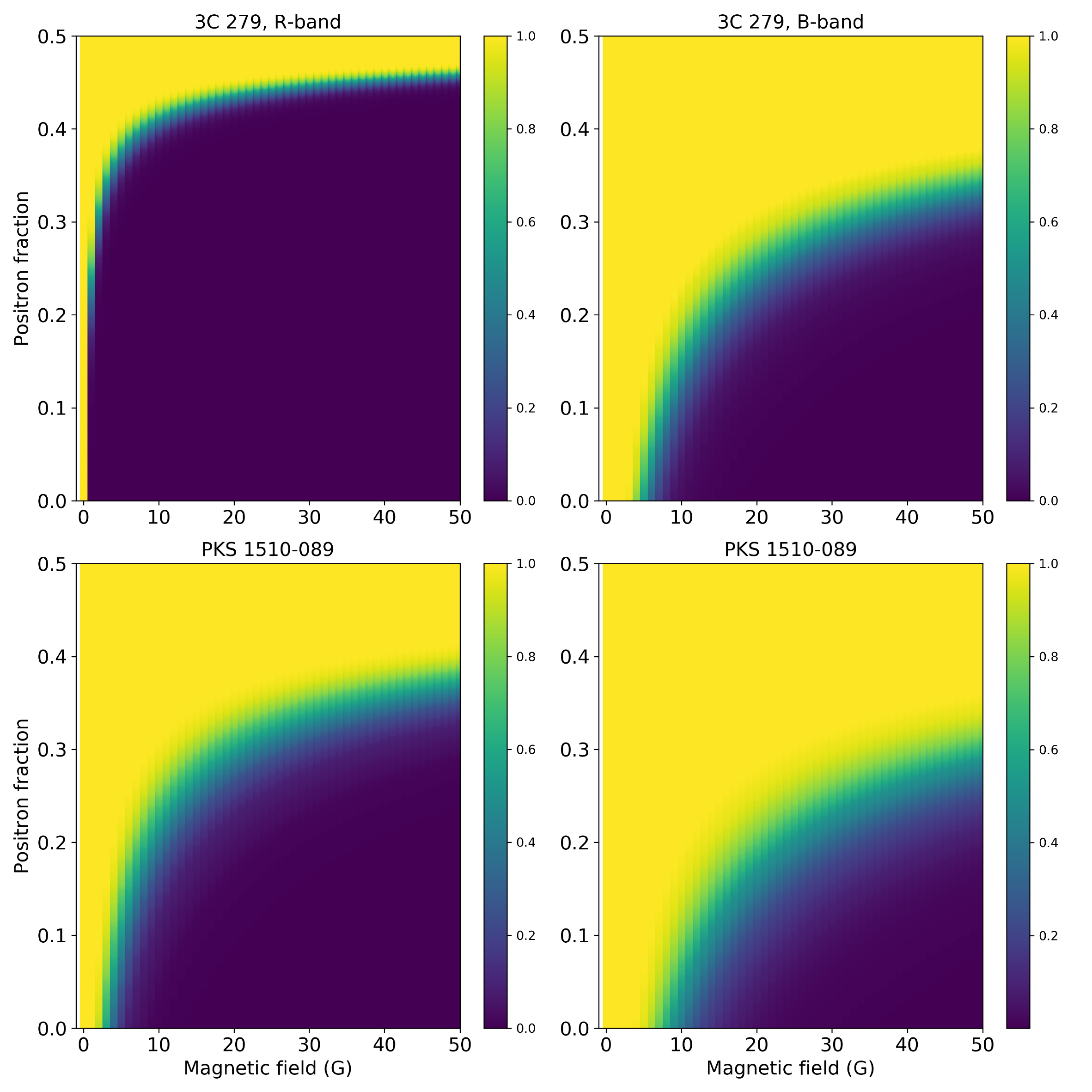}}
\caption{Constraints on the magnetic field strength and jet composition for both sources in different observing sessions. The top row shows the results for 3C~279 for the R-band (left) and B-band (right). The bottom row shows the results for PKS~1510$-$089 at MJD~59312.645 (left) and MJD~59313.645 (right). The colorbar shows the fraction of accepted random $\delta$ draws given $B,~f$.}
\label{plt:constraints}
\end{figure*}

Synchrotron radiation is characterized by elliptical polarization, with a small fractional degree of intrinsic CP, $\Pi_c \propto \nu^{-1/2}$  \citep{Legg1968, Melrose1971}. In the case of a pure pair plasma (electron-positron) the circular component cancels leaving only the linear component of the polarization. Here we 
assume that the jet plasma is made up of an electron-positron-proton (pair-proton) composition. In order for the jets to be accelerated to the high observed velocities (Lorentz factor $\Gamma>30$) at pc-scales \citep{Liodakis2017,Liodakis2018-II}, would require at least some population of ``cold'' protons \citep{Phinney1987,Sikora1996,Sikora1996-II}, however, those do not contribute significantly to the emission. In this scenario, we expect at least some degree of intrinsic CP that depends on the LP, jet composition, strength of magnetic field and $\Gamma$ of the bulk flow. For a power-law emitting particle distribution with index $\sim(2-2.3)$, a perpendicular intrinsic magnetic field and an optically-thin jet at a $\theta\approx1/\Gamma$ angle to the observer, we can relate the (intrinsic) magnetic field strength to the degree of CP as,
\begin{equation}
B \approx 2\times 10^7 \left( \frac{\nu_{\rm obs}}{10^{15}~Hz}\right)   \left( \frac{0.71}{\Pi_l}\right)^{2}  \left( \frac{1}{\Gamma^3(1-2f)^2}\right)  \,     \Pi_c^2,
\label{Eq:bfield}
\end{equation}
where, $B$ is in Gauss, $\nu_{\rm obs}$ the observing frequency, $\Pi_l$ the linear polarization degree, $f$ the fraction of positrons, and $\Pi_c$ 
the circular polarization degree \citep{Rieger2005}. The term $(0.71/\Pi_l)$ describes the uniformity of the magnetic field which affects LP and CP equally \citep{Jones1977}. In the $\theta\approx1/\Gamma$ regime, $\Gamma\approx\delta$, where $\delta$ is the Doppler factor, $\delta=1/[\Gamma(1-\beta\cos\theta)]$. In \cite{Liodakis2021} we found that 3C~279 is decelerating in the GHz range, with $\delta$ peaking above 37~GHz and below 100~GHz. Therefore we use $\delta=18.3\pm1.9$ derived at 43~GHz \citep{Jorstad2017}. PKS~1510$-$089 does not show $\delta$ variations across the GHz range \citep{Liodakis2018-II,Liodakis2021}, hence for consistency we use $\delta=35.3\pm4.6$ also found at 43~GHz.  We discuss the choice of $\delta$ below.

To fully explore the parameter space of $B,~f,~\delta$ we uniformly draw random values in the range of [1,100] Gauss for $B$ and [0,0.5) for $f$, where $f=0$ is for a pure electron-proton plasma and $f=0.5$ is for a pure electron-positron plasma. We also draw a random value for $\delta$ from a Gaussian distribution with mean and standard deviation $\delta$ and $\sigma_\delta$ respectively. Using the observed $\Pi_l$, observing frequency and Eq. \ref{Eq:bfield} we can estimate $\Pi_c$ which we compare to the observed value. We repeat this process $10^6$ times and record all $B,~f$ pairs that satisfy the observed 2$\sigma$ upper limits for $\Pi_c$ for each of the observing dates and sources. Figure \ref{plt:constraints} shows the $f$ versus $B$ plane with the colorbar showing the fraction of accepted random $\delta$ draws given $B,~f$. We have omitted the PKS~1510$-$089 observation at MJD~59315.631 since the low observed $\Pi_l$ and high $\Pi_c$ upper limit do not provide any constraints on $B$ and $f$.  We find a range of values for both parameters with the high $B$-field -- low $f$ parameter space to be excluded for both sources. Our best constraints come from MJD~59316.519 for 3C~279 and MJD~59312.645 for PKS~1510$-$089 (Table \ref{tab:Table1}, Fig \ref{plt:constraints} top and bottom left panels).

\section{Discussion \& Conclusions}

Here we presented measurements of the optical circular polarization degree for two well-studied blazars, namely 3C~279 and PKS~1510$-$089. This is the first time an attempt was made to measure CP in PKS~1510$-$089. Although our measurements can only be considered as upper limits, they still provide meaningful constrains on the magnetic field and jet composition. Generally, we find that emission models requiring high magnetic field strengths and a low positron composition can be excluded. Leptonic models typically require $B < 5 G$ while high $B$-fields ($>30G$) are often required in proton-synchrotron models (see \citealp{Cerruti2020} for a review). This is due to their low radiative efficiency which depends on $B$ \citep{Sikora2009}. Specifically, SED hadronic modeling estimates for the magnetic field strength in 3C~279 and PKS~1510$-$089 are in the range of 50 - 150~G and 10 - 50~G, respectively  (e.g., \citealp{Boettcher2013,Diltz2015,Paliya2018}). Such values only allow for large positron fractions in 3C~279 ($f>0.43$). In \cite{Liodakis2020} we estimated the minimum magnetic field strength in the context of a proton-synchrotron model to be about 43~G for 3C~279 and 9~G for PKS~1510$-$089. For 3C~279 this limits the range of compositions to $0.43\leq{f}<0.5$. This is consistent with \cite{Wardle1998} who found through radio CP that the jet in 3C~279 is composed mainly from an electron-positron plasma. For PKS~1510$-$089 the low $B$ estimate does not allow us to place any constraints on its positron fraction. However, we note that proton-synchrotron models typically assume a pure proton-electron plasma. A lower proton fraction would require even higher magnetic field strengths. In addition, in order for proton synchrotron to dominate the steady state high-energy emission, both sources require more than two orders of magnitude higher luminosity than the available Blandford-Znajek power of the jets, even in the MAD regime \citep{Liodakis2020}. Lepto-hadronic models or models with subdominant hadronic components are viable alternatives for low magnetic field strengths ($B<10~G$, e.g.,  \citealp{Petropoulou2015,Mastichiadis2013,Gao2019}). Such field strength levels are often found through radio observations \citep{Pushkarev2012}. If the jets contain a significant fraction of protons, our results would be in favor of the aforementioned models, noting however, that their energetic requirements can often be greater than that of proton-synchrotron models.

Throughout this work we have used as our limiting $\Pi_c$ value the 2$\sigma$ upper limit of each individual observation. In the case of 3C~279, using a 3$\sigma$ limit has only a mild quantitative effect and does not change any of our conclusions. However, in the case of PKS~1510$-$089 for $\delta<20$ we cannot exclude models requiring high $B$-fields and low positron fractions. One of our main assumptions is that all of the non-detected CP signal is intrinsic to the jet. If the observed CP limits include both an intrinsic and extrinsic contribution, our constraints on the $f$, $B$ would improve. We have also assumed the $\delta$ of the optical emitting regions to be similar to the ones derived in radio. This is likely to be true for a number of blazars that do not show  $\delta$ variations even at very high radio frequencies \citep{Liodakis2021}.  While the $\delta$ estimate for 3C~279 is modest, PKS~1510$-$089 has a rather high value. However, we note that the $\sigma_\delta = 1.9$ for 3C~279 in  \cite{Jorstad2017} is optimistic. The spread of the reported proper motions is rather large. In our study \citep{Blinov2021-II} we found that the ratios of $\delta$ for individual components can change up to a factor $\sim 5$. Assuming a larger Doppler factor will only tighten the constraints on $B$ and $f$. Therefore, targeting high linearly polarized -- high-$\delta$ blazars, or during outbursts when $\delta$ could be amplified \citep{Larionov2010,Raiteri2017-II,Uemura2017,Liodakis2020-II}, would allow us to further constrain the B-field and jet composition in blazars.

\section*{Acknowledgements}
We thank the anonymous referee for constructive comments that helped improve this work. We also thank Maria Petropoulou for useful discussions and Sergey Savchenko for providing alerts on high polarization states of blazars.  D.B. acknowledges support from the European Research Council (ERC) under the European Union Horizon 2020 research and innovation programme under the grant agreement No 771282. I.L thanks the University of Crete for their hospitality while this paper was written.

\section*{Data Availability}

The data underlying this article will be shared on reasonable request to the corresponding author.








\bibliographystyle{mnras}
\bibliography{bibliography} 

\bsp	
\label{lastpage}
\end{document}